\documentclass[aps,nofootinbib, prl, amsmath, floats, floatfix, twocolumn,superscriptaddress, showpacs,reprint]{revtex4}

\usepackage{graphicx}
\usepackage{amsmath,amssymb}
\usepackage{amsfonts}
\usepackage{xspace} 
\usepackage[usenames]{color}
\usepackage{dcolumn}
\usepackage{bm}
\usepackage{mathrsfs}
\usepackage[colorlinks=true]{hyperref}
\usepackage[all]{hypcap} 
 \usepackage[utf8]{inputenc} 

\usepackage{ulem}
\definecolor {darkgreen}{rgb}{0.2,0.7,0.2}

\newcommand{\eq}{\begin{equation}}
\newcommand{\be}{\begin{equation}}
\newcommand{\eeq}{\end{equation}}
\newcommand{\ee}{\end{equation}}

\newcommand{\GW}{{\mbox{\tiny GW}}}
\newcommand{\ppE}{{\mbox{\tiny ppE}}}

\begin{document}

\title{Theory-Agnostic Constraints on Black-Hole Dipole Radiation \\ with Multi-Band Gravitational-Wave Astrophysics}

\author{Enrico Barausse}
\affiliation{Sorbonne Universit\'es, UPMC Univ Paris 6 \& CNRS, UMR 7095, Institut d'Astrophysique de Paris, 98 bis bd Arago, 75014 Paris, France}
\author{Nicol\'as Yunes}
\affiliation{eXtreme Gravity Institute, Department of Physics, Montana State University, Bozeman, MT 59717, USA.}
\author{Katie Chamberlain}
\affiliation{eXtreme Gravity Institute, Department of Physics, Montana State University, Bozeman, MT 59717, USA.}

\begin{abstract}
The aLIGO detection of the black-hole binary GW150914 opened a new era for probing extreme gravity. Many gravity theories predict the emission of dipole gravitational radiation by binaries. This is excluded to high accuracy in binary pulsars, but entire classes of theories predict this effect predominantly (or only) in binaries involving black holes. Joint observations of GW150914-like systems by aLIGO and eLISA will improve bounds on dipole emission from black-hole binaries by six orders of magnitude relative to current constraints, provided that eLISA is not dramatically descoped. 
\end{abstract}

\pacs{04.30.-w, 04.80.Cc, 04.80.Nn}
\date{\today \hspace{0.2truecm}}

\maketitle

\textit{Introduction.} The advanced LIGO (aLIGO)~\cite{ligo}  observation of gravitational waves (GWs) from GW150914~\cite{GW150914} heralds a new era in astrophysics in which GWs will probe the universe in a new and complementary way to traditional telescopes~\cite{Yunes:2013dva}. The GWs detected were emitted in the late inspiral, merger, and ringdown of a black-hole (BH) binary with masses $36^{+5}_{-4} M_\odot$ and $29^{+4}_{-4} M_\odot$ at a redshift $z=0.09^{+0.03}_{-0.04}$~\cite{GW150914}.  While the existence of GWs had been demonstrated indirectly by their backreaction on the orbital evolution of binary pulsars~\cite{damour-taylor}, GW150914 represents their first direct detection,  and also provides the most convincing evidence to date of the very existence of BHs. 

With several new GW detectors coming online soon, GW150914 is just the beginning. Advanced Virgo~\cite{virgo} is expected to  start science runs in 2016, while KAGRA~\cite{kagra} is under construction and LIGO India~\cite{ligo-india} has been approved. Pulsar timing arrays~\cite{epta,nanograv,ppta,2010CQGra..27h4013H} are already observing at nHz frequencies, and ESA's evolving Laser Interferometer Space Antenna (eLISA)~\cite{Seoane:2013qna} will target mHz frequencies; these frequency bands are complementary to the $10$--$10^3$ Hz range probed by terrestrial detectors. Some GW sources will likely have electromagnetic or neutrino/cosmic-ray counterparts, which could provide additional information on their nature, formation, and environment. 

GW150914-like sources will not only be visible in the aLIGO band, but also in the mHz eLISA band, thus opening the prospect for \emph{multi-band} GW astronomy~\cite{Sesana:2016ljz,AmaroSeoane:2009ui}). 
Indeed, \cite{Sesana:2016ljz} predicts that anywhere from  a few to thousands of BH binaries with total masses $\sim 50$--$100 M_\odot$
will be detected by eLISA in a 5 year mission. These systems inspiral in the eLISA band for years
before chirping out of it and re-appearing in the aLIGO band to merge typically a few weeks later. Observations with eLISA will allow measurements of system parameters before detection with aLIGO. This may help localize the source in the sky in
a timely/more accurate manner, thus increasing the chances of finding an electromagnetic counterpart,
and will also allow for the prediction of coalescence time in the aLIGO band weeks/months in advance, with errors $\lesssim 10\,$s~\cite{Sesana:2016ljz}.

This exciting prospect, however, depends heavily on whether or not the eLISA design is dramatically descoped, a topic currently under investigation by ESA (see e.g.~\cite{2016PhRvD..93b4003K,2015arXiv151206239C,2016arXiv160107112T,Porter:2015bea} for investigations of the impact of different designs on eLISA science). Indeed, as highlighted in~\cite{Sesana:2016ljz}, cost-saving measures, such as the decrease in the number of laser links from six to four, or shortening the mission duration or the length of the interferometer arms, may dangerously impact eLISA's capability to resolve GW150914-like binaries. In this Letter, we show that an excessive descope of the  eLISA configuration -- one that makes the detection of GW150914-like BH binaries impossible -- will miss out on an incredible opportunity to perform a generic test of gravitational theories: to constrain the existence of BH dipole gravitational radiation to an accuracy that surpasses current constraints by six orders of magnitude.

\textit{Motion and GW emission in GR extensions.} 
Gravity theories that extend GR typically affect the motion of bodies and GW emission. Though details of these changes to the dynamics depend on the specific theory, some common traits can be identified. In most theories, the gravitational field is described by a spin-2 metric tensor field, and by additional fields (see e.g.~\cite{Berti:2015itd}). The interaction between matter and the new fields may give rise to ``fifth forces'',
both conservative and dissipative. The latter can be thought of as energy-momentum exchanges between matter and the new fields, i.e.~the stress-energy tensor of matter is not generally conserved. 
For weakly gravitating bodies or in regimes of weak gravitational fields (like on Earth or in the Solar System), modifications to the dynamics are excluded to high confidence by particle-physics and gravitational experiments~\cite{will-living}. However, if the additional 
fields do not couple to matter at tree level, their effect on the motion will be suppressed, i.e.~the ``weak'' equivalence principle -- the universality of free fall in weak-gravity regimes -- will be satisfied, and these experimental tests will be passed. 

The motion of \emph{strongly gravitating} bodies, such as neutron stars (NSs) and BHs, can more easily deviate from the GR expectation in modified gravity theories.  Indeed, an effective coupling between the extra fields and matter, even if suppressed at tree level, typically re-appears at higher perturbative orders. This is because the extra fields generally couple non-minimally to the metric, which in turn is coupled to matter via gravity. Therefore, when gravity is strong, the non-minimal coupling causes 
the emergence of (effective) fifth forces and energy-momentum exchanges between matter and the extra fields, thus leading to deviations from the universality of free fall~\cite{eardley,Nordtvedt:1968qr}. These 
are referred to as  violations of the ``strong'' equivalence principle, or ``N\"ordtvedt effect''.

A modification of the motion of strongly gravitating bodies will leave an imprint in the GWs these bodies emit. In GR, GW emission is predominantly quadrupolar, as monopole and dipole emission are forbidden by the conservation of the matter stress-energy tensor. In modified gravity, however, the matter stress-energy tensor is generally not conserved due to the N\"ordtvedt effect, thus allowing monopole and dipole emission~\cite{will-living}. Dipole radiation, in particular, is the dominant effect for quasi-circular binary systems, although its actual presence and magnitude generally depend on the nature of the binary components and the modified theory of gravity in question. Besides dipole radiation, conservative modifications to the dynamics (e.g. to the binary's binding energy/Hamiltonian) may also be present, but they are typically subdominant as they enter at higher post-Newtonian (PN) order\footnote{The PN approximation solves the field equations perturbatively in the ratio $(v/c)$, where $v$ is the binary's relative velocity. Terms suppressed by $(v/c)^{2n}$ relative to leading order are said to be of nPN order.}~\cite{Yagi:2013qpa,PhysRevD.89.084067,Barausse:2014tra}.  

To understand how dipole radiation comes about, consider one of the simplest GR extensions. In ``scalar-tensor'' (ST) theories of the Fierz, Jordan, Brans and Dicke (FJBD) type~\cite{Fierz:1956zz,Jordan:1959eg,Brans:1961sx}, the gravitational interaction is mediated by the metric and by a gravitational scalar. The latter has a standard kinetic term in the action (up to a field redefinition), is minimally coupled to matter, and is directly coupled to the Ricci scalar. Because of the standard kinetic term, the scalar obeys the Klein-Gordon equation, with a source (due to the coupling to the Ricci scalar in the action) that depends on the matter stress-energy. Therefore, the scalar is not excited in globally vacuum spacetimes, and can only be non-constant because of non-trivial boundary or initial conditions (e.g.~if the scalar field is not initially uniform, in which case it undergoes a transient evolution before settling to a constant~\cite{Healy:2011ef}, or if cosmological or non-asymptotically-flat boundary conditions are imposed~\cite{Horbatsch:2011ye,Berti:2013gfa}). Therefore, BH spacetimes (isolated or binary) generally do not excite a scalar field (i.e.~have ``no hair'') and do not emit dipole radiation in these theories~\cite{Yunes:2011aa,Mirshekari:2013vb}. 

Nevertheless, FJBD-like ST theories predict that dipole emission should be present in binaries involving at least one NS. This has been historically very important, because binary pulsar observations constrain deviations of the orbital period decay away from the GR prediction to high accuracy. For example, the double binary pulsar PSR J0737-3039~\cite{manchester,lyne} constrains $\delta\equiv |(\dot{P}/{{P}})_{\rm non\,GR}-(\dot{P}/{{P}})_{\rm GR}|/(\dot{P}/{{P}})_{\rm GR} \lesssim 10^{-2}$~\cite{stairs,kramer-double-pulsar,Yunes:2010qb}, while the binary pulsar J1141-6545~\cite{bhat,2002PhRvD..66b4040G} constrains $\delta \lesssim 6 \times 10^{-4}$. These observations place very stringent constraints on several gravitational theories, including FJBD-like ST ones. 

To derive a precise bound on gravitational dipole emission with binary-pulsar observations, let us parametrize a dipole flux correction as
\begin{equation}\label{flux}
\dot{E}_{\rm GW}=\dot{E}_{\rm GR} \left[1+B \left(\frac{G m}{r_{12} c^2}\right)^{-1}\right]
\end{equation}
where $\dot{E}_{\rm GR}$ is the GR GW flux (given at leading order by the quadrupole formula), $m$ and $r_{12}$ are the binary's total mass and orbital separation, and $B$ is a theory-dependent parameter regulating the strength of the dipole term (e.g.~in FJBD-like ST theories, $B=5 (\Delta \alpha)^2/96$, where $\Delta \alpha$ is the difference between the scalar charges of the two bodies~\cite{damour_esposito_farese,Mirshekari:2013vb}). Dipole emission is enhanced (relative to quadrupolar emission) by a factor $(Gm/r_{12} c^2)^{-1}$, i.e.~a -1PN effect dominating over the GR prediction at large separations (or low frequencies). Since $\dot{P}/{P}=-(3/2) \dot{E}_{\rm GW}/|E_{\rm b}|$, $E_{\rm b}$ being the Newtonian binding energy of the binary,  one obtains $\delta=|\dot{E}_{\rm GW}/\dot{E}_{\rm GR}-1|=|B| ({G m}/{r_{12}c^2})^{-1}\lesssim 10^{-2}$. This leads to the  approximate bound $|B|\lesssim 6 \times 10^{-8}$ with PSR 0737-3039~\cite{Yunes:2010qb} and $|B|\lesssim 2 \times 10^{-9}$ with PSR J1141--6545. Other binary pulsars lead to similar bounds.

These bounds place stringent constraints on several theories that predict dipole GW emission in the inspiral of binaries involving at least one NS, e.g.~numerous FJBD-like ST theories (especially those predicting spontaneous scalarization for isolated NSs~\cite{damour_esposito_farese,ST0,freire,Wex:2014nva}), Lorentz-violating gravity~\cite{PhysRevD.89.084067,Yagi:2013qpa}, or theories with a MOND-like phenomenology~\cite{Bonetti:2015oda}. Binary pulsars, however, are less efficient at testing theories where dipole radiation activates late in the inspiral -- e.g.~certain FJBD-like ST theories where NSs do not spontaneously scalarize in isolation, but undergo ``dynamical scalarization'' in close binaries~\cite{ST1,ST2,Sampson:2014qqa,ST3,Taniguchi:2014fqa,Ponce:2014hha} -- or theories that predict dipole emission predominantly (or only) in BH binaries.

Let us focus on the latter case. ST theories more general than FJBD-like ones couple the scalar field not only to the Ricci scalar, but also to more general curvature invariants such as the Gauss-Bonnet invariant or the Pontryagin density~\cite{Yagi:2015oca,Yagi:2016jml}. In these cases, the equation for the scalar has a source that depends not only on the matter stress-energy (like in FJBD-like theories) but also on matter-independent curvature terms. The former typically induces dipole radiation in NS binaries and follows from coupling the scalar field to the Ricci scalar in the action, while the latter do not vanish in vacuum, may induce dipole radiation in BH binaries, and follow from coupling the scalar to more general curvature invariants. Therefore, bounds on dipole emission from binary pulsars and  BH binaries are \textit{complementary}, as they constrain different couplings in the action. As an extreme example, one can select these couplings to eliminate dipole emission in binary pulsars altogether, while retaining dipole emission in BH binaries. This is the case in shift-symmetric dilatonic Gauss-Bonnet gravity~\cite{Barausse:2015wia,Yagi:2015oca}, where the scalar interacts with the curvature only via a linear coupling to the Gauss-Bonnet invariant in the action\footnote{In these theories dipole emission from NS binaries vanishes due to both \textit{(i)} shift symmetry and \textit{(ii)} NS-NS spacetimes being simply connected~\cite{Yagi:2015oca,Barausse:2015wia}. Assumption \textit{(ii)} is not valid for BH spacetimes, which is why BH binaries emit dipole radiation while NS-NS binaries do not, although the latter will present deviations from GR at higher PN orders.}.

A similar situation is expected in theories with vector fields (e.g.~Lorentz violating gravity and theories with a MOND phenomenology) or tensor fields (i.e.~bimetric gravity theories), where the contribution of the extra fields does not vanish in vacuum. This causes BHs to possess extra ``hairs'' besides the GR ones (mass and spin)~\cite{Barausse:2011pu,Barausse:2013nwa,Barausse:2012ny,Barausse:2012qh,Barausse:2015frm,Brito:2013xaa,Babichev:2015xha}, leading to dipole emission from BH binaries, at least in principle\footnote{A rigorous proof of the existence of BH dipole radiation in these theories is not yet available, as GW emission in the presence of extra vector and tensor modes is quite involved, c.f. the calculation  in \cite{PhysRevD.89.084067} for NS binaries in Lorentz-violating gravity.}. Finally, as mentioned above, even within FJBD-like ST theories, dipole BH emission might arise from non-trivial (e.g. cosmological) boundary conditions. In light of all this, it makes sense to constrain dipole radiation without theoretical bias.

Currently, the most stringent constraint on vacuum dipole radiation follows from the orbital decay rate of the A0620-00 low-mass X-ray binary (LMXB), a main sequence star in orbit around a BH. By assuming that the observed orbital decay is consistent with GR, one obtains the $1\sigma$ constraint $|B| < 1.9 \times 10^{-3}$~\cite{kent-LMXB}. This bound is sensitive to systematics in the astrophysical model, e.g. the BH is accreting from the star, hence the orbital decay rate depends on the mass transfer rate and on the angular momentum carried away by stellar winds, which are not known a priori. 
 
\textit{Projected constraints on dipole GW emission from BH binaries.} Since aLIGO only sees BH binaries near merger, the GW dipole term is subdominant in the aLIGO band, and thus can only be weakly constrained~\cite{LIGO-ppE}. However, if a GW150914-like BH binary is first observed by eLISA and then by aLIGO, dipole emission can be constrained with exquisite precision. 

Let us provide an approximate physical argument for why this is so. If GR is correct, the GW150914 binary was emitting at a GW frequency $f_{\GW,i} \approx 0.016$ Hz (in the middle of the eLISA band) 5 years before merger. If eLISA had been in operation in its best configuration (i.e. the ``classic LISA'' N2A5M5L6 configuration of \cite{2016PhRvD..93b4003K}), it would have measured the system parameters with outstanding accuracy, predicting in particular the time at which the system would have merged in the aLIGO band to within $10 \,$s. However, if the GR flux is modified by a dipole term as in Eq.~\eqref{flux}, the binary will coalesce earlier, because dipole emission will shed additional energy and angular momentum. To qualitatively assess this effect, consider the evolution of the GW frequency $f_{\GW}$ under Eq.~\eqref{flux}: ${{\rm d} f_{\GW}}/{{\rm d} t}=- ({{\rm d} f_{\GW}}/{{\rm d} E_b}) \dot{E}_{\GW}$. One can compute the time $\Delta t$ needed by GW150914 to evolve from $f_{\GW,i}$ to $f_{\GW,m}\approx 132$ Hz by integrating $({{\rm d} f_{\GW}}/{{\rm d} t})^{-1}$. By requiring that the difference $|(\Delta t)_{GR}-(\Delta t)_{\rm non\, GR}|$ be less than an uncertainty of $10\,$s on the merger time, a joint eLISA-aLIGO observation of GW150914 would roughly constrain $|B|\lesssim 10^{-10}$. 

To improve this rough estimate, we perform a Fisher-matrix analysis to obtain bounds on $B$. We use the (quasi-circular, non-spin-precessing) PhenomB inspiral-merger-ringdown waveform of~\cite{Ajith:2007kx,Ajith:2009bn}, with the addition of a single parametrized post-Einsteinian~\cite{PPE,Chatziioannou:2012rf,Barausse:2014tra} inspiral phase term at $-1$ PN order ($\Psi_{\ppE} = \beta (\pi {\cal{M}} f)^{b/3}$, with $\beta = -(3/224) \eta^{2/5} B$ and $b=-7$, where $\eta = m_{1} m_{2}/m^{2}$, $m = m_{1} + m_{2}$ is the  total mass and ${\cal{M}} = m \eta^{3/5}$ is the chirp mass). The parameters of this model are then $(A,\phi_{c},t_{c},{\cal{M}},\eta,\chi_{\rm eff},B)$, where $A$ is an overall amplitude, $(\phi_{c},t_{c})$ are the phase and time of coalescence and $\chi_{\rm eff}$ is an effective spin parameter. 

We take the aLIGO noise curve at design sensitivity from~\cite{LIGOSn}, and that at the time of the GW150914 detection from~\cite{noise-data}. For eLISA, we use sky-averaged, six-link sensitivity curves, since GW150914-like events will be considerably more difficult to resolve with four links~\cite{Sesana:2016ljz}. We consider a 5 year mission, and allow multiple options for the arm length (1, 2, or 5 Gm, i.e.~A1, A2, A5) and the low frequency noise (N2 for the expected LISA Pathfinder performance,  N1 for a noise ten  times  worse)~\cite{2016PhRvD..93b4003K}. We assume that the observation is simultaneously done by two instruments (either the two independent eLISA interferometers, or the aLIGO Hanford and Livingston sites). To combine aLIGO and eLISA results, we add the Fisher matrices and then invert the sum to obtain the variance-covariance matrix~\cite{Berti:2004bd,2009arXiv0906.0664H}.\footnote{This is a good approximation if the signal in the two detectors is phase-connected, i.e. if the chirp mass
is measured by eLISA with sufficient accuracy so as to account for all the cycles between the two bands~\cite{neil}. We have verified that this
is indeed the case.}

We explore the projected bounds on $B$
with several BH binaries.
For GW150914-like systems, we consider total masses $m=(50,80,100) M_\odot$ (as well as $m=65 M_{\odot}$ for the actual GW150914 event),
a mass ratio $q\equiv m_{1}/m_{2}=0.8$, dimensionless spin parameters $(\chi_1,\chi_2)=(0.4,0.3)$, and a luminosity distance $d_L=400$ Mpc (i.e.~redshift $z\sim 0.085$). 
For massive BH binaries, we consider $m=(10^4,10^5,10^6) M_\odot$, 
with large $(\chi_1,\chi_2)=(0.9,0.8)$ spins, mass ratios $q=(0.3,0.8)$, and $d_L=(16,48)$ Gpc (i.e.~$z\sim 2$ and $5$). We also consider extreme/intermediate mass-ratio inspirals (EMRIs/IMRIs), 
for which we take
individual masses $(10,10^{5}) M_{\odot}$, $(10,10^{4}) M_{\odot}$, $(10^{2},10^{5}) M_{\odot}$, $(10^{3},10^{5}) M_{\odot}$,
$d_L=(1,5)$ Gpc [$z\sim(0.2,0.8)$],  and spins $(\chi_{1},\chi_{2})=(0.5,0.8)$ in all cases.

\begin{figure}[tb]
\includegraphics[clip=true,width=\columnwidth{}]{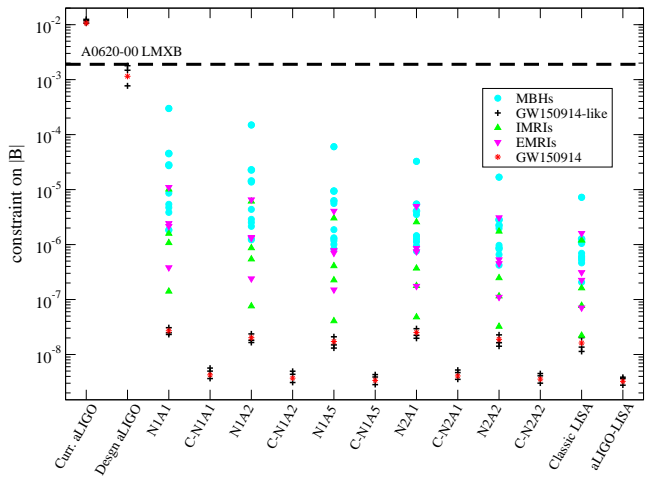}
\caption{\label{figure1} $1\sigma$ constraints on the BH dipole flux parameter $B$ from various sources -- GW150914-like BH binaries (stars and plusses), massive BH binaries (filled circles) and EMRIs/IMRIs (filled triangles) -- as a function of the instrument -- aLIGO at the time of the GW150914 event (Curr.~aLIGO), aLIGO at design sensitivity (Desgn aLIGO), various six-link eLISA configurations (NxAx) with target/pessimistic low-frequency noise (N2/N1) and 1 Gm/2 Gm arms (A1/A2), a Classic LISA design with six links, 5-Gm arms and target-level low-frequency noise, and joint observations by design aLIGO and eLISA (C-NxAx), or by  design aLIGO and Classic LISA (aLIGO-LISA). For comparison, we also include the current constraint on vacuum dipole radiation from LMXB A0620-00~\cite{kent-LMXB}. The combined aLIGO-eLISA observation of GW150914-like sources leads to the most stringent constraints, which are six orders of magnitude stronger than current bounds. 
}
\end{figure}

Figure~\ref{figure1} summarizes our projected $1 \sigma$ bounds on $B$ and compares them to the existing ones. eLISA observations of GW150914-like systems lead to constraints typically five orders of magnitude stronger than the current A0620-00 constraints, and six orders of magnitude stronger than current aLIGO constraints. This is because when these binaries produce GWs in the eLISA band, they are widely separated and thus emit dipole radiation abundantly. For example, 5 years prior to merger (while in the eLISA band), the GW150914 binary had an orbital velocity of $0.025 \; c$. When exiting the eLISA band at $\sim 0.1$ Hz, the velocity was $0.048 \;c$, which increased to $0.22 \;c$ upon entering the aLIGO band at $\sim 10$Hz. Notice that eLISA is sensitive to the very low-velocity/early inspiral, where not only does dipole radiation dominate over quadrupole radiation, but any systematics due to the PN approximation are negligible (unlike for events including the merger). We have confirmed this by repeating the Fisher analysis with two waveform models (PhenomB and PhenomD~\cite{TheLIGOScientific:2016src,LIGO-ppE}), which lead to very similar constraints on dipole radiation, since the models are almost indistinguishable in the early inspiral (see also~\cite{LIGO-ppE}).

Combined eLISA and design-aLIGO observations lead to constraints $\sim 10$ times better than eLISA observations alone. If design aLIGO is upgraded, e.g. to one of the LIGO Explorer designs with a ten-fold increase in sensitivity at $100$Hz~\cite{ScienceWhite}, these combined constraints would become $\sim 10^{2}$ times better than eLISA observations alone.
Note that the eLISA/design-aLIGO combined constraints are roughly one order of magnitude worse than the approximate calculation presented earlier, because the latter does not account for correlations between parameters. We have indeed verified that the bound from the Fisher analysis becomes stronger and approaches the approximate estimate if we assume that all parameters except $\beta$ (or equivalently $B$) are known exactly, i.e. if we assume that the variance on $\beta$ is simply given by the inverse of the corresponding diagonal Fisher matrix entry. 

eLISA observations of massive BH binaries and EMRIs/IMRIs would also constrain BH dipole radiation, although bounds are typically weaker. How strong these constraints are depends on the orbital separation (or relative velocity) of these binaries when they emit in the eLISA band. For example, eLISA will be sensitive only to the late inspiral and merger-ringdown of very massive BH binaries, which is why these lead to weaker constraints in Fig.~\ref{figure1}.

One may wonder whether the projected constraints discussed above are robust, since gravity modifications inducing dipole emission, if present, will typically change the GW model not only at -1PN order in the waveform phase, but also at higher PN orders. However, as shown explicitly in~\cite{LIGO-ppE}, at least in FJBD theory, these higher-order PN corrections only affect a Fisher analysis like ours by at most $10\%$. The addition of many terms in the GW phase at multiple PN orders is not only unnecessary, but actually counterproductive as it dilutes the ability to extract information from the signal (as pointed out in~\cite{Sampson:2013jpa,cornish-PPE} and verified in~\cite{TheLIGOScientific:2016src}).

Finally, our constraints apply to theories where 
the gravity modifications are not completely screened on the scale of BH binaries. Indeed, 
 there is no proposed mechanism that can completely screen gravity 
modifications at small scales in \textit{dynamical} situations such as those of interest here (see e.g.~\cite{Bonetti:2015oda,Jimenez:2015bwa,deRham:2012fw,deRham:2012fg} and discussion
in \cite{Barausse:2015wia}). Similarly, we implicitly assume that modified gravity effects do not
appear suddenly and non-perturbatively at specific energy scales/resonances (see e.g.~\cite{Cardoso:2011xi,Yunes:2011aa,ST1,ST2,Sampson:2014qqa,ST3,Taniguchi:2014fqa,Ponce:2014hha}).

\textit{Conclusion.} The realization that GW150914-like BH binaries could be multi-band GW-astronomy targets for aLIGO and eLISA opens a unique door to study various physical mechanisms in extreme-gravity regimes. The generation of dipole radiation (and other potential effects) can be tested by such joint eLISA-aLIGO observations with unprecedented precision. The exciting prospect of multi-band observation, however, will only materialize if the eLISA design is not excessively descoped.
 
\begin{acknowledgments}
\emph{Acknowledgments}. We would like to thank Kent Yagi for helpful discussions and code comparisons, Alberto Sesana for useful discussions about the detectability of GW150914-like sources by eLISA, and Emanuele Berti and Vitor Cardoso for reading a preliminary version of this manuscript and providing useful and insightful comments. Special thanks to Neil Cornish for insightful comments on how to combine aLIGO and eLISA observations. EB acknowledges support from the European Union's Seventh Framework Programme (FP7/PEOPLE-2011-CIG) through the Marie Curie Career Integration Grant GALFORMBHS PCIG11-GA-2012-321608, and from the H2020-MSCA-RISE-2015 Grant No. StronGrHEP-690904. NY and KC acknowledge support from the NSF CAREER Grant PHY-1250636. 
\end{acknowledgments}

\bibliography{master}

\begin{thebibliography}{80}
\expandafter\ifx\csname natexlab\endcsname\relax\def\natexlab#1{#1}\fi
\expandafter\ifx\csname bibnamefont\endcsname\relax
  \def\bibnamefont#1{#1}\fi
\expandafter\ifx\csname bibfnamefont\endcsname\relax
  \def\bibfnamefont#1{#1}\fi
\expandafter\ifx\csname citenamefont\endcsname\relax
  \def\citenamefont#1{#1}\fi
\expandafter\ifx\csname url\endcsname\relax
  \def\url#1{\texttt{#1}}\fi
\expandafter\ifx\csname urlprefix\endcsname\relax\def\urlprefix{URL }\fi
\providecommand{\bibinfo}[2]{#2}
\providecommand{\eprint}[2][]{\url{#2}}

\bibitem[{lig({\natexlab{a}})}]{ligo}
\emph{\bibinfo{title}{{LIGO}}}, \bibinfo{note}{{\tt www.ligo.caltech.edu}}.

\bibitem[{\citenamefont{Abbott et~al.}(2016{\natexlab{a}})}]{GW150914}
\bibinfo{author}{\bibfnamefont{B.~P.} \bibnamefont{Abbott}}
  \bibnamefont{et~al.} (\bibinfo{collaboration}{Virgo, LIGO Scientific}),
  \bibinfo{journal}{Phys. Rev. Lett.} \textbf{\bibinfo{volume}{116}},
  \bibinfo{pages}{061102} (\bibinfo{year}{2016}{\natexlab{a}}),
  \eprint{1602.03837}.

\bibitem[{\citenamefont{Yunes and Siemens}(2013)}]{Yunes:2013dva}
\bibinfo{author}{\bibfnamefont{N.}~\bibnamefont{Yunes}} \bibnamefont{and}
  \bibinfo{author}{\bibfnamefont{X.}~\bibnamefont{Siemens}},
  \bibinfo{journal}{Living Reviews in Relativity} \textbf{\bibinfo{volume}{16}}
  (\bibinfo{year}{2013}),
  \urlprefix\url{http://www.livingreviews.org/lrr-2013-9}.

\bibitem[{\citenamefont{Damour and Taylor}(1992)}]{damour-taylor}
\bibinfo{author}{\bibfnamefont{T.}~\bibnamefont{Damour}} \bibnamefont{and}
  \bibinfo{author}{\bibfnamefont{J.~H.} \bibnamefont{Taylor}},
  \bibinfo{journal}{Phys. Rev. D} \textbf{\bibinfo{volume}{45}},
  \bibinfo{pages}{1840} (\bibinfo{year}{1992}).

\bibitem[{vir()}]{virgo}
\emph{\bibinfo{title}{{VIRGO}}}, \bibinfo{note}{{\tt www.virgo.infn.it}}.

\bibitem[{kag()}]{kagra}
\emph{\bibinfo{title}{Kagra}}, \bibinfo{note}{{\tt
  http://gwcenter.icrr.u-tokyo.ac.jp/en/}}.

\bibitem[{lig({\natexlab{b}})}]{ligo-india}
\emph{\bibinfo{title}{{LIGO India}}}, \bibinfo{note}{{\tt www.gw-indigo.org}}.

\bibitem[{ept()}]{epta}
\emph{\bibinfo{title}{{EPTA}}}, \bibinfo{note}{{\tt www.epta.eu.org}}.

\bibitem[{nan()}]{nanograv}
\emph{\bibinfo{title}{{NANOGrav}}}, \bibinfo{note}{{\tt nanograv.org}}.

\bibitem[{ppt()}]{ppta}
\emph{\bibinfo{title}{{PPTA}}}, \bibinfo{note}{{\tt
  www.atnf.csiro.au/research/pulsar/ppta}}.

\bibitem[{\citenamefont{{Hobbs} et~al.}(2010)}]{2010CQGra..27h4013H}
\bibinfo{author}{\bibfnamefont{G.}~\bibnamefont{{Hobbs}}} \bibnamefont{et~al.},
  \bibinfo{journal}{Classical and Quantum Gravity}
  \textbf{\bibinfo{volume}{27}}, \bibinfo{eid}{084013} (\bibinfo{year}{2010}),
  \bibinfo{note}{arXiv:0911.5206}.

\bibitem[{\citenamefont{Seoane et~al.}(2013)}]{Seoane:2013qna}
\bibinfo{author}{\bibfnamefont{P.~A.} \bibnamefont{Seoane}}
  \bibnamefont{et~al.} (\bibinfo{collaboration}{eLISA Collaboration})
  (\bibinfo{year}{2013}), \eprint{1305.5720}.

\bibitem[{\citenamefont{Sesana}(2016)}]{Sesana:2016ljz}
\bibinfo{author}{\bibfnamefont{A.}~\bibnamefont{Sesana}}
  (\bibinfo{year}{2016}), \eprint{1602.06951}.

\bibitem[{\citenamefont{Amaro-Seoane and
  Santamaria}(2010)}]{AmaroSeoane:2009ui}
\bibinfo{author}{\bibfnamefont{P.}~\bibnamefont{Amaro-Seoane}}
  \bibnamefont{and}
  \bibinfo{author}{\bibfnamefont{L.}~\bibnamefont{Santamaria}},
  \bibinfo{journal}{Astrophys. J.} \textbf{\bibinfo{volume}{722}},
  \bibinfo{pages}{1197} (\bibinfo{year}{2010}), \eprint{0910.0254}.

\bibitem[{\citenamefont{{Klein} et~al.}(2016)\citenamefont{{Klein}, {Barausse},
  {Sesana}, {Petiteau}, {Berti}, {Babak}, {Gair}, {Aoudia}, {Hinder}, {Ohme}
  et~al.}}]{2016PhRvD..93b4003K}
\bibinfo{author}{\bibfnamefont{A.}~\bibnamefont{{Klein}}},
  \bibinfo{author}{\bibfnamefont{E.}~\bibnamefont{{Barausse}}},
  \bibinfo{author}{\bibfnamefont{A.}~\bibnamefont{{Sesana}}},
  \bibinfo{author}{\bibfnamefont{A.}~\bibnamefont{{Petiteau}}},
  \bibinfo{author}{\bibfnamefont{E.}~\bibnamefont{{Berti}}},
  \bibinfo{author}{\bibfnamefont{S.}~\bibnamefont{{Babak}}},
  \bibinfo{author}{\bibfnamefont{J.}~\bibnamefont{{Gair}}},
  \bibinfo{author}{\bibfnamefont{S.}~\bibnamefont{{Aoudia}}},
  \bibinfo{author}{\bibfnamefont{I.}~\bibnamefont{{Hinder}}},
  \bibinfo{author}{\bibfnamefont{F.}~\bibnamefont{{Ohme}}},
  \bibnamefont{et~al.}, \bibinfo{journal}{\prd} \textbf{\bibinfo{volume}{93}},
  \bibinfo{eid}{024003} (\bibinfo{year}{2016}), \eprint{1511.05581}.

\bibitem[{\citenamefont{{Caprini} et~al.}(2015)\citenamefont{{Caprini},
  {Hindmarsh}, {Huber}, {Konstandin}, {Kozaczuk}, {Nardini}, {No}, {Petiteau},
  {Schwaller}, {Servant} et~al.}}]{2015arXiv151206239C}
\bibinfo{author}{\bibfnamefont{C.}~\bibnamefont{{Caprini}}},
  \bibinfo{author}{\bibfnamefont{M.}~\bibnamefont{{Hindmarsh}}},
  \bibinfo{author}{\bibfnamefont{S.}~\bibnamefont{{Huber}}},
  \bibinfo{author}{\bibfnamefont{T.}~\bibnamefont{{Konstandin}}},
  \bibinfo{author}{\bibfnamefont{J.}~\bibnamefont{{Kozaczuk}}},
  \bibinfo{author}{\bibfnamefont{G.}~\bibnamefont{{Nardini}}},
  \bibinfo{author}{\bibfnamefont{J.~M.} \bibnamefont{{No}}},
  \bibinfo{author}{\bibfnamefont{A.}~\bibnamefont{{Petiteau}}},
  \bibinfo{author}{\bibfnamefont{P.}~\bibnamefont{{Schwaller}}},
  \bibinfo{author}{\bibfnamefont{G.}~\bibnamefont{{Servant}}},
  \bibnamefont{et~al.}, \bibinfo{journal}{ArXiv e-prints}
  (\bibinfo{year}{2015}), \eprint{1512.06239}.

\bibitem[{\citenamefont{{Tamanini} et~al.}(2016)\citenamefont{{Tamanini},
  {Caprini}, {Barausse}, {Sesana}, {Klein}, and
  {Petiteau}}}]{2016arXiv160107112T}
\bibinfo{author}{\bibfnamefont{N.}~\bibnamefont{{Tamanini}}},
  \bibinfo{author}{\bibfnamefont{C.}~\bibnamefont{{Caprini}}},
  \bibinfo{author}{\bibfnamefont{E.}~\bibnamefont{{Barausse}}},
  \bibinfo{author}{\bibfnamefont{A.}~\bibnamefont{{Sesana}}},
  \bibinfo{author}{\bibfnamefont{A.}~\bibnamefont{{Klein}}}, \bibnamefont{and}
  \bibinfo{author}{\bibfnamefont{A.}~\bibnamefont{{Petiteau}}},
  \bibinfo{journal}{ArXiv e-prints}  (\bibinfo{year}{2016}),
  \eprint{1601.07112}.

\bibitem[{\citenamefont{Porter}(2015)}]{Porter:2015bea}
\bibinfo{author}{\bibfnamefont{E.~K.} \bibnamefont{Porter}},
  \bibinfo{journal}{Phys. Rev.} \textbf{\bibinfo{volume}{D92}},
  \bibinfo{pages}{064001} (\bibinfo{year}{2015}), \eprint{1505.08058}.

\bibitem[{\citenamefont{Berti et~al.}(2015)}]{Berti:2015itd}
\bibinfo{author}{\bibfnamefont{E.}~\bibnamefont{Berti}} \bibnamefont{et~al.}
  (\bibinfo{year}{2015}), \eprint{1501.07274}.

\bibitem[{\citenamefont{Will}(2006)}]{will-living}
\bibinfo{author}{\bibfnamefont{C.~M.} \bibnamefont{Will}},
  \bibinfo{journal}{Living Reviews in Relativity} \textbf{\bibinfo{volume}{9}}
  (\bibinfo{year}{2006}),
  \urlprefix\url{http://www.livingreviews.org/lrr-2006-3}.

\bibitem[{\citenamefont{{Eardley}}(1975)}]{eardley}
\bibinfo{author}{\bibfnamefont{D.~M.} \bibnamefont{{Eardley}}},
  \bibinfo{journal}{Astrophys. J. Lett.} \textbf{\bibinfo{volume}{196}},
  \bibinfo{pages}{L59} (\bibinfo{year}{1975}).

\bibitem[{\citenamefont{Nordtvedt}(1968)}]{Nordtvedt:1968qr}
\bibinfo{author}{\bibfnamefont{K.}~\bibnamefont{Nordtvedt}},
  \bibinfo{journal}{Phys. Rev.} \textbf{\bibinfo{volume}{169}},
  \bibinfo{pages}{1014} (\bibinfo{year}{1968}).

\bibitem[{\citenamefont{Yagi et~al.}(2014{\natexlab{a}})\citenamefont{Yagi,
  Blas, Yunes, and Barausse}}]{Yagi:2013qpa}
\bibinfo{author}{\bibfnamefont{K.}~\bibnamefont{Yagi}},
  \bibinfo{author}{\bibfnamefont{D.}~\bibnamefont{Blas}},
  \bibinfo{author}{\bibfnamefont{N.}~\bibnamefont{Yunes}}, \bibnamefont{and}
  \bibinfo{author}{\bibfnamefont{E.}~\bibnamefont{Barausse}},
  \bibinfo{journal}{Phys. Rev. Lett.} \textbf{\bibinfo{volume}{112}},
  \bibinfo{pages}{161101} (\bibinfo{year}{2014}{\natexlab{a}}).

\bibitem[{\citenamefont{Yagi et~al.}(2014{\natexlab{b}})\citenamefont{Yagi,
  Blas, Barausse, and Yunes}}]{PhysRevD.89.084067}
\bibinfo{author}{\bibfnamefont{K.}~\bibnamefont{Yagi}},
  \bibinfo{author}{\bibfnamefont{D.}~\bibnamefont{Blas}},
  \bibinfo{author}{\bibfnamefont{E.}~\bibnamefont{Barausse}}, \bibnamefont{and}
  \bibinfo{author}{\bibfnamefont{N.}~\bibnamefont{Yunes}},
  \bibinfo{journal}{Phys. Rev. D} \textbf{\bibinfo{volume}{89}},
  \bibinfo{pages}{084067} (\bibinfo{year}{2014}{\natexlab{b}}).

\bibitem[{\citenamefont{Barausse et~al.}(2014)\citenamefont{Barausse, Cardoso,
  and Pani}}]{Barausse:2014tra}
\bibinfo{author}{\bibfnamefont{E.}~\bibnamefont{Barausse}},
  \bibinfo{author}{\bibfnamefont{V.}~\bibnamefont{Cardoso}}, \bibnamefont{and}
  \bibinfo{author}{\bibfnamefont{P.}~\bibnamefont{Pani}},
  \bibinfo{journal}{Phys. Rev.} \textbf{\bibinfo{volume}{D89}},
  \bibinfo{pages}{104059} (\bibinfo{year}{2014}), \eprint{1404.7149}.

\bibitem[{\citenamefont{Fierz}(1956)}]{Fierz:1956zz}
\bibinfo{author}{\bibfnamefont{M.}~\bibnamefont{Fierz}},
  \bibinfo{journal}{Helv. Phys. Acta} \textbf{\bibinfo{volume}{29}},
  \bibinfo{pages}{128} (\bibinfo{year}{1956}).

\bibitem[{\citenamefont{Jordan}(1959)}]{Jordan:1959eg}
\bibinfo{author}{\bibfnamefont{P.}~\bibnamefont{Jordan}}, \bibinfo{journal}{Z.
  Phys.} \textbf{\bibinfo{volume}{157}}, \bibinfo{pages}{112}
  (\bibinfo{year}{1959}).

\bibitem[{\citenamefont{Brans and Dicke}(1961)}]{Brans:1961sx}
\bibinfo{author}{\bibfnamefont{C.}~\bibnamefont{Brans}} \bibnamefont{and}
  \bibinfo{author}{\bibfnamefont{R.~H.} \bibnamefont{Dicke}},
  \bibinfo{journal}{Phys. Rev.} \textbf{\bibinfo{volume}{124}},
  \bibinfo{pages}{925} (\bibinfo{year}{1961}).

\bibitem[{\citenamefont{Healy et~al.}(2012)\citenamefont{Healy, Bode, Haas,
  Pazos, Laguna, Shoemaker, and Yunes}}]{Healy:2011ef}
\bibinfo{author}{\bibfnamefont{J.}~\bibnamefont{Healy}},
  \bibinfo{author}{\bibfnamefont{T.}~\bibnamefont{Bode}},
  \bibinfo{author}{\bibfnamefont{R.}~\bibnamefont{Haas}},
  \bibinfo{author}{\bibfnamefont{E.}~\bibnamefont{Pazos}},
  \bibinfo{author}{\bibfnamefont{P.}~\bibnamefont{Laguna}},
  \bibinfo{author}{\bibfnamefont{D.~M.} \bibnamefont{Shoemaker}},
  \bibnamefont{and} \bibinfo{author}{\bibfnamefont{N.}~\bibnamefont{Yunes}},
  \bibinfo{journal}{Class. Quant. Grav.} \textbf{\bibinfo{volume}{29}},
  \bibinfo{pages}{232002} (\bibinfo{year}{2012}), \eprint{1112.3928}.

\bibitem[{\citenamefont{Horbatsch and Burgess}(2012)}]{Horbatsch:2011ye}
\bibinfo{author}{\bibfnamefont{M.~W.} \bibnamefont{Horbatsch}}
  \bibnamefont{and} \bibinfo{author}{\bibfnamefont{C.~P.}
  \bibnamefont{Burgess}}, \bibinfo{journal}{JCAP}
  \textbf{\bibinfo{volume}{1205}}, \bibinfo{pages}{010} (\bibinfo{year}{2012}),
  \eprint{1111.4009}.

\bibitem[{\citenamefont{Berti et~al.}(2013)\citenamefont{Berti, Cardoso,
  Gualtieri, Horbatsch, and Sperhake}}]{Berti:2013gfa}
\bibinfo{author}{\bibfnamefont{E.}~\bibnamefont{Berti}},
  \bibinfo{author}{\bibfnamefont{V.}~\bibnamefont{Cardoso}},
  \bibinfo{author}{\bibfnamefont{L.}~\bibnamefont{Gualtieri}},
  \bibinfo{author}{\bibfnamefont{M.}~\bibnamefont{Horbatsch}},
  \bibnamefont{and} \bibinfo{author}{\bibfnamefont{U.}~\bibnamefont{Sperhake}},
  \bibinfo{journal}{Phys. Rev.} \textbf{\bibinfo{volume}{D87}},
  \bibinfo{pages}{124020} (\bibinfo{year}{2013}), \eprint{1304.2836}.

\bibitem[{\citenamefont{Yunes et~al.}(2012)\citenamefont{Yunes, Pani, and
  Cardoso}}]{Yunes:2011aa}
\bibinfo{author}{\bibfnamefont{N.}~\bibnamefont{Yunes}},
  \bibinfo{author}{\bibfnamefont{P.}~\bibnamefont{Pani}}, \bibnamefont{and}
  \bibinfo{author}{\bibfnamefont{V.}~\bibnamefont{Cardoso}},
  \bibinfo{journal}{Phys. Rev.} \textbf{\bibinfo{volume}{D85}},
  \bibinfo{pages}{102003} (\bibinfo{year}{2012}), \eprint{1112.3351}.

\bibitem[{\citenamefont{Mirshekari and Will}(2013)}]{Mirshekari:2013vb}
\bibinfo{author}{\bibfnamefont{S.}~\bibnamefont{Mirshekari}} \bibnamefont{and}
  \bibinfo{author}{\bibfnamefont{C.~M.} \bibnamefont{Will}},
  \bibinfo{journal}{Phys. Rev.} \textbf{\bibinfo{volume}{D87}},
  \bibinfo{pages}{084070} (\bibinfo{year}{2013}), \eprint{1301.4680}.

\bibitem[{\citenamefont{{Manchester} et~al.}(2005)\citenamefont{{Manchester},
  {Kramer}, {Possenti}, {Lyne}, {Burgay}, {Stairs}, {Hotan}, {McLaughlin},
  {Lorimer}, {Hobbs} et~al.}}]{manchester}
\bibinfo{author}{\bibfnamefont{R.~N.} \bibnamefont{{Manchester}}},
  \bibinfo{author}{\bibfnamefont{M.}~\bibnamefont{{Kramer}}},
  \bibinfo{author}{\bibfnamefont{A.}~\bibnamefont{{Possenti}}},
  \bibinfo{author}{\bibfnamefont{A.~G.} \bibnamefont{{Lyne}}},
  \bibinfo{author}{\bibfnamefont{M.}~\bibnamefont{{Burgay}}},
  \bibinfo{author}{\bibfnamefont{I.~H.} \bibnamefont{{Stairs}}},
  \bibinfo{author}{\bibfnamefont{A.~W.} \bibnamefont{{Hotan}}},
  \bibinfo{author}{\bibfnamefont{M.~A.} \bibnamefont{{McLaughlin}}},
  \bibinfo{author}{\bibfnamefont{D.~R.} \bibnamefont{{Lorimer}}},
  \bibinfo{author}{\bibfnamefont{G.~B.} \bibnamefont{{Hobbs}}},
  \bibnamefont{et~al.}, \bibinfo{journal}{Astrophys. J. Lett.}
  \textbf{\bibinfo{volume}{621}}, \bibinfo{pages}{L49} (\bibinfo{year}{2005}).

\bibitem[{\citenamefont{{Lyne} et~al.}(2004)\citenamefont{{Lyne}, {Burgay},
  {Kramer}, {Possenti}, {Manchester}, {Camilo}, {McLaughlin}, {Lorimer},
  {D'Amico}, {Joshi} et~al.}}]{lyne}
\bibinfo{author}{\bibfnamefont{A.~G.} \bibnamefont{{Lyne}}},
  \bibinfo{author}{\bibfnamefont{M.}~\bibnamefont{{Burgay}}},
  \bibinfo{author}{\bibfnamefont{M.}~\bibnamefont{{Kramer}}},
  \bibinfo{author}{\bibfnamefont{A.}~\bibnamefont{{Possenti}}},
  \bibinfo{author}{\bibfnamefont{R.~N.} \bibnamefont{{Manchester}}},
  \bibinfo{author}{\bibfnamefont{F.}~\bibnamefont{{Camilo}}},
  \bibinfo{author}{\bibfnamefont{M.~A.} \bibnamefont{{McLaughlin}}},
  \bibinfo{author}{\bibfnamefont{D.~R.} \bibnamefont{{Lorimer}}},
  \bibinfo{author}{\bibfnamefont{N.}~\bibnamefont{{D'Amico}}},
  \bibinfo{author}{\bibfnamefont{B.~C.} \bibnamefont{{Joshi}}},
  \bibnamefont{et~al.}, \bibinfo{journal}{Science}
  \textbf{\bibinfo{volume}{303}}, \bibinfo{pages}{1153} (\bibinfo{year}{2004}).

\bibitem[{\citenamefont{Stairs}(2003)}]{stairs}
\bibinfo{author}{\bibfnamefont{I.~H.} \bibnamefont{Stairs}},
  \bibinfo{journal}{Living Rev.Rel.} \textbf{\bibinfo{volume}{6}},
  \bibinfo{pages}{5} (\bibinfo{year}{2003}).

\bibitem[{\citenamefont{Kramer et~al.}(2006)\citenamefont{Kramer, Stairs,
  Manchester, McLaughlin, Lyne et~al.}}]{kramer-double-pulsar}
\bibinfo{author}{\bibfnamefont{M.}~\bibnamefont{Kramer}},
  \bibinfo{author}{\bibfnamefont{I.~H.} \bibnamefont{Stairs}},
  \bibinfo{author}{\bibfnamefont{R.}~\bibnamefont{Manchester}},
  \bibinfo{author}{\bibfnamefont{M.}~\bibnamefont{McLaughlin}},
  \bibinfo{author}{\bibfnamefont{A.}~\bibnamefont{Lyne}}, \bibnamefont{et~al.},
  \bibinfo{journal}{Science} \textbf{\bibinfo{volume}{314}},
  \bibinfo{pages}{97} (\bibinfo{year}{2006}).

\bibitem[{\citenamefont{Yunes and Hughes}(2010)}]{Yunes:2010qb}
\bibinfo{author}{\bibfnamefont{N.}~\bibnamefont{Yunes}} \bibnamefont{and}
  \bibinfo{author}{\bibfnamefont{S.~A.} \bibnamefont{Hughes}},
  \bibinfo{journal}{Phys. Rev.} \textbf{\bibinfo{volume}{D82}},
  \bibinfo{pages}{082002} (\bibinfo{year}{2010}), \eprint{1007.1995}.

\bibitem[{\citenamefont{Bhat et~al.}(2008)\citenamefont{Bhat, Bailes, and
  Verbiest}}]{bhat}
\bibinfo{author}{\bibfnamefont{N.~R.} \bibnamefont{Bhat}},
  \bibinfo{author}{\bibfnamefont{M.}~\bibnamefont{Bailes}}, \bibnamefont{and}
  \bibinfo{author}{\bibfnamefont{J.~P.} \bibnamefont{Verbiest}},
  \bibinfo{journal}{Phys.Rev.} \textbf{\bibinfo{volume}{D77}},
  \bibinfo{pages}{124017} (\bibinfo{year}{2008}).

\bibitem[{\citenamefont{{G{\'e}rard} and {Wiaux}}(2002)}]{2002PhRvD..66b4040G}
\bibinfo{author}{\bibfnamefont{J.-M.} \bibnamefont{{G{\'e}rard}}}
  \bibnamefont{and} \bibinfo{author}{\bibfnamefont{Y.}~\bibnamefont{{Wiaux}}},
  \bibinfo{journal}{\prd} \textbf{\bibinfo{volume}{66}}, \bibinfo{eid}{024040}
  (\bibinfo{year}{2002}), \eprint{gr-qc/0109062}.

\bibitem[{\citenamefont{{Damour} and
  {Esposito-Farese}}(1992)}]{damour_esposito_farese}
\bibinfo{author}{\bibfnamefont{T.}~\bibnamefont{{Damour}}} \bibnamefont{and}
  \bibinfo{author}{\bibfnamefont{G.}~\bibnamefont{{Esposito-Farese}}},
  \bibinfo{journal}{Classical and Quantum Gravity}
  \textbf{\bibinfo{volume}{9}}, \bibinfo{pages}{2093} (\bibinfo{year}{1992}).

\bibitem[{\citenamefont{{Damour} and {Esposito-Farese}}(1993)}]{ST0}
\bibinfo{author}{\bibfnamefont{T.}~\bibnamefont{{Damour}}} \bibnamefont{and}
  \bibinfo{author}{\bibfnamefont{G.}~\bibnamefont{{Esposito-Farese}}},
  \bibinfo{journal}{Phys. Rev. Lett.} \textbf{\bibinfo{volume}{70}},
  \bibinfo{pages}{2220} (\bibinfo{year}{1993}).

\bibitem[{\citenamefont{Freire et~al.}(2012)\citenamefont{Freire, Wex,
  Esposito-Farese, Verbiest, Bailes et~al.}}]{freire}
\bibinfo{author}{\bibfnamefont{P.~C.} \bibnamefont{Freire}},
  \bibinfo{author}{\bibfnamefont{N.}~\bibnamefont{Wex}},
  \bibinfo{author}{\bibfnamefont{G.}~\bibnamefont{Esposito-Farese}},
  \bibinfo{author}{\bibfnamefont{J.~P.} \bibnamefont{Verbiest}},
  \bibinfo{author}{\bibfnamefont{M.}~\bibnamefont{Bailes}},
  \bibnamefont{et~al.}, \bibinfo{journal}{Mon. Not. Roy. Astron. Soc.}
  \textbf{\bibinfo{volume}{423}}, \bibinfo{pages}{3328} (\bibinfo{year}{2012}).

\bibitem[{\citenamefont{Wex}(2014)}]{Wex:2014nva}
\bibinfo{author}{\bibfnamefont{N.}~\bibnamefont{Wex}} (\bibinfo{year}{2014}),
  \eprint{1402.5594}.

\bibitem[{\citenamefont{Bonetti and Barausse}(2015)}]{Bonetti:2015oda}
\bibinfo{author}{\bibfnamefont{M.}~\bibnamefont{Bonetti}} \bibnamefont{and}
  \bibinfo{author}{\bibfnamefont{E.}~\bibnamefont{Barausse}},
  \bibinfo{journal}{Phys. Rev.} \textbf{\bibinfo{volume}{D91}},
  \bibinfo{pages}{084053} (\bibinfo{year}{2015}), \bibinfo{note}{[Erratum:
  Phys. Rev.D93,029901(2016)]}, \eprint{1502.05554}.

\bibitem[{\citenamefont{Barausse et~al.}(2013)\citenamefont{Barausse,
  Palenzuela, Ponce, and Lehner}}]{ST1}
\bibinfo{author}{\bibfnamefont{E.}~\bibnamefont{Barausse}},
  \bibinfo{author}{\bibfnamefont{C.}~\bibnamefont{Palenzuela}},
  \bibinfo{author}{\bibfnamefont{M.}~\bibnamefont{Ponce}}, \bibnamefont{and}
  \bibinfo{author}{\bibfnamefont{L.}~\bibnamefont{Lehner}},
  \bibinfo{journal}{Phys. Rev. D} \textbf{\bibinfo{volume}{87}},
  \bibinfo{pages}{081506} (\bibinfo{year}{2013}).

\bibitem[{\citenamefont{Palenzuela et~al.}(2014)\citenamefont{Palenzuela,
  Barausse, Ponce, and Lehner}}]{ST2}
\bibinfo{author}{\bibfnamefont{C.}~\bibnamefont{Palenzuela}},
  \bibinfo{author}{\bibfnamefont{E.}~\bibnamefont{Barausse}},
  \bibinfo{author}{\bibfnamefont{M.}~\bibnamefont{Ponce}}, \bibnamefont{and}
  \bibinfo{author}{\bibfnamefont{L.}~\bibnamefont{Lehner}},
  \bibinfo{journal}{Phys. Rev. D} \textbf{\bibinfo{volume}{89}},
  \bibinfo{pages}{044024} (\bibinfo{year}{2014}).

\bibitem[{\citenamefont{Sampson
  et~al.}(2014{\natexlab{a}})\citenamefont{Sampson, Yunes, Cornish, Ponce,
  Barausse, Klein, Palenzuela, and Lehner}}]{Sampson:2014qqa}
\bibinfo{author}{\bibfnamefont{L.}~\bibnamefont{Sampson}},
  \bibinfo{author}{\bibfnamefont{N.}~\bibnamefont{Yunes}},
  \bibinfo{author}{\bibfnamefont{N.}~\bibnamefont{Cornish}},
  \bibinfo{author}{\bibfnamefont{M.}~\bibnamefont{Ponce}},
  \bibinfo{author}{\bibfnamefont{E.}~\bibnamefont{Barausse}},
  \bibinfo{author}{\bibfnamefont{A.}~\bibnamefont{Klein}},
  \bibinfo{author}{\bibfnamefont{C.}~\bibnamefont{Palenzuela}},
  \bibnamefont{and} \bibinfo{author}{\bibfnamefont{L.}~\bibnamefont{Lehner}},
  \bibinfo{journal}{Phys. Rev. D} \textbf{\bibinfo{volume}{90}},
  \bibinfo{pages}{124091} (\bibinfo{year}{2014}{\natexlab{a}}).

\bibitem[{\citenamefont{Shibata et~al.}(2014)\citenamefont{Shibata, Taniguchi,
  Okawa, and Buonanno}}]{ST3}
\bibinfo{author}{\bibfnamefont{M.}~\bibnamefont{Shibata}},
  \bibinfo{author}{\bibfnamefont{K.}~\bibnamefont{Taniguchi}},
  \bibinfo{author}{\bibfnamefont{H.}~\bibnamefont{Okawa}}, \bibnamefont{and}
  \bibinfo{author}{\bibfnamefont{A.}~\bibnamefont{Buonanno}},
  \bibinfo{journal}{Phys. Rev. D} \textbf{\bibinfo{volume}{89}},
  \bibinfo{pages}{084005} (\bibinfo{year}{2014}).

\bibitem[{\citenamefont{Taniguchi et~al.}(2015)\citenamefont{Taniguchi,
  Shibata, and Buonanno}}]{Taniguchi:2014fqa}
\bibinfo{author}{\bibfnamefont{K.}~\bibnamefont{Taniguchi}},
  \bibinfo{author}{\bibfnamefont{M.}~\bibnamefont{Shibata}}, \bibnamefont{and}
  \bibinfo{author}{\bibfnamefont{A.}~\bibnamefont{Buonanno}},
  \bibinfo{journal}{Phys. Rev.} \textbf{\bibinfo{volume}{D91}},
  \bibinfo{pages}{024033} (\bibinfo{year}{2015}), \eprint{1410.0738}.

\bibitem[{\citenamefont{Ponce et~al.}(2015)\citenamefont{Ponce, Palenzuela,
  Barausse, and Lehner}}]{Ponce:2014hha}
\bibinfo{author}{\bibfnamefont{M.}~\bibnamefont{Ponce}},
  \bibinfo{author}{\bibfnamefont{C.}~\bibnamefont{Palenzuela}},
  \bibinfo{author}{\bibfnamefont{E.}~\bibnamefont{Barausse}}, \bibnamefont{and}
  \bibinfo{author}{\bibfnamefont{L.}~\bibnamefont{Lehner}},
  \bibinfo{journal}{Phys. Rev.} \textbf{\bibinfo{volume}{D91}},
  \bibinfo{pages}{084038} (\bibinfo{year}{2015}), \eprint{1410.0638}.

\bibitem[{\citenamefont{Yagi et~al.}(2016)\citenamefont{Yagi, Stein, and
  Yunes}}]{Yagi:2015oca}
\bibinfo{author}{\bibfnamefont{K.}~\bibnamefont{Yagi}},
  \bibinfo{author}{\bibfnamefont{L.~C.} \bibnamefont{Stein}}, \bibnamefont{and}
  \bibinfo{author}{\bibfnamefont{N.}~\bibnamefont{Yunes}},
  \bibinfo{journal}{Phys. Rev.} \textbf{\bibinfo{volume}{D93}},
  \bibinfo{pages}{024010} (\bibinfo{year}{2016}), \eprint{1510.02152}.

\bibitem[{\citenamefont{Yagi and Stein}(2016)}]{Yagi:2016jml}
\bibinfo{author}{\bibfnamefont{K.}~\bibnamefont{Yagi}} \bibnamefont{and}
  \bibinfo{author}{\bibfnamefont{L.~C.} \bibnamefont{Stein}},
  \bibinfo{journal}{Class. Quant. Grav.} \textbf{\bibinfo{volume}{33}},
  \bibinfo{pages}{054001} (\bibinfo{year}{2016}), \eprint{1602.02413}.

\bibitem[{\citenamefont{Barausse and Yagi}(2015)}]{Barausse:2015wia}
\bibinfo{author}{\bibfnamefont{E.}~\bibnamefont{Barausse}} \bibnamefont{and}
  \bibinfo{author}{\bibfnamefont{K.}~\bibnamefont{Yagi}},
  \bibinfo{journal}{Phys. Rev. Lett.} \textbf{\bibinfo{volume}{115}},
  \bibinfo{pages}{211105} (\bibinfo{year}{2015}), \eprint{1509.04539}.

\bibitem[{\citenamefont{Barausse et~al.}(2011)\citenamefont{Barausse, Jacobson,
  and Sotiriou}}]{Barausse:2011pu}
\bibinfo{author}{\bibfnamefont{E.}~\bibnamefont{Barausse}},
  \bibinfo{author}{\bibfnamefont{T.}~\bibnamefont{Jacobson}}, \bibnamefont{and}
  \bibinfo{author}{\bibfnamefont{T.~P.} \bibnamefont{Sotiriou}},
  \bibinfo{journal}{Phys.Rev.} \textbf{\bibinfo{volume}{D83}},
  \bibinfo{pages}{124043} (\bibinfo{year}{2011}).

\bibitem[{\citenamefont{Barausse and
  Sotiriou}(2013{\natexlab{a}})}]{Barausse:2013nwa}
\bibinfo{author}{\bibfnamefont{E.}~\bibnamefont{Barausse}} \bibnamefont{and}
  \bibinfo{author}{\bibfnamefont{T.~P.} \bibnamefont{Sotiriou}},
  \bibinfo{journal}{Class. Quant. Grav.} \textbf{\bibinfo{volume}{30}},
  \bibinfo{pages}{244010} (\bibinfo{year}{2013}{\natexlab{a}}),
  \eprint{1307.3359}.

\bibitem[{\citenamefont{Barausse and Sotiriou}(2012)}]{Barausse:2012ny}
\bibinfo{author}{\bibfnamefont{E.}~\bibnamefont{Barausse}} \bibnamefont{and}
  \bibinfo{author}{\bibfnamefont{T.~P.} \bibnamefont{Sotiriou}},
  \bibinfo{journal}{Phys. Rev. Lett.} \textbf{\bibinfo{volume}{109}},
  \bibinfo{pages}{181101} (\bibinfo{year}{2012}), \bibinfo{note}{erratum-ibid.\
  {\bf 110}, 039902 (2013)}.

\bibitem[{\citenamefont{Barausse and
  Sotiriou}(2013{\natexlab{b}})}]{Barausse:2012qh}
\bibinfo{author}{\bibfnamefont{E.}~\bibnamefont{Barausse}} \bibnamefont{and}
  \bibinfo{author}{\bibfnamefont{T.~P.} \bibnamefont{Sotiriou}},
  \bibinfo{journal}{Phys.Rev.} \textbf{\bibinfo{volume}{D87}},
  \bibinfo{pages}{087504} (\bibinfo{year}{2013}{\natexlab{b}}).

\bibitem[{\citenamefont{Barausse et~al.}(2016)\citenamefont{Barausse, Sotiriou,
  and Vega}}]{Barausse:2015frm}
\bibinfo{author}{\bibfnamefont{E.}~\bibnamefont{Barausse}},
  \bibinfo{author}{\bibfnamefont{T.~P.} \bibnamefont{Sotiriou}},
  \bibnamefont{and} \bibinfo{author}{\bibfnamefont{I.}~\bibnamefont{Vega}},
  \bibinfo{journal}{Phys. Rev.} \textbf{\bibinfo{volume}{D93}},
  \bibinfo{pages}{044044} (\bibinfo{year}{2016}), \eprint{1512.05894}.

\bibitem[{\citenamefont{Brito et~al.}(2013)\citenamefont{Brito, Cardoso, and
  Pani}}]{Brito:2013xaa}
\bibinfo{author}{\bibfnamefont{R.}~\bibnamefont{Brito}},
  \bibinfo{author}{\bibfnamefont{V.}~\bibnamefont{Cardoso}}, \bibnamefont{and}
  \bibinfo{author}{\bibfnamefont{P.}~\bibnamefont{Pani}},
  \bibinfo{journal}{Phys. Rev.} \textbf{\bibinfo{volume}{D88}},
  \bibinfo{pages}{064006} (\bibinfo{year}{2013}), \eprint{1309.0818}.

\bibitem[{\citenamefont{Babichev and Brito}(2015)}]{Babichev:2015xha}
\bibinfo{author}{\bibfnamefont{E.}~\bibnamefont{Babichev}} \bibnamefont{and}
  \bibinfo{author}{\bibfnamefont{R.}~\bibnamefont{Brito}},
  \bibinfo{journal}{Class. Quant. Grav.} \textbf{\bibinfo{volume}{32}},
  \bibinfo{pages}{154001} (\bibinfo{year}{2015}), \eprint{1503.07529}.

\bibitem[{\citenamefont{Yagi}(2012)}]{kent-LMXB}
\bibinfo{author}{\bibfnamefont{K.}~\bibnamefont{Yagi}}, \bibinfo{journal}{Phys.
  Rev. D} \textbf{\bibinfo{volume}{86}}, \bibinfo{pages}{081504}
  (\bibinfo{year}{2012}).

\bibitem[{\citenamefont{Yunes et~al.}()\citenamefont{Yunes, Yagi, and
  Pretorius}}]{LIGO-ppE}
\bibinfo{author}{\bibfnamefont{N.}~\bibnamefont{Yunes}},
  \bibinfo{author}{\bibfnamefont{K.}~\bibnamefont{Yagi}}, \bibnamefont{and}
  \bibinfo{author}{\bibfnamefont{F.}~\bibnamefont{Pretorius}},
  \emph{\bibinfo{title}{in preparation}}.

\bibitem[{\citenamefont{Ajith et~al.}(2008)}]{Ajith:2007kx}
\bibinfo{author}{\bibfnamefont{P.}~\bibnamefont{Ajith}} \bibnamefont{et~al.},
  \bibinfo{journal}{Phys. Rev.} \textbf{\bibinfo{volume}{D77}},
  \bibinfo{pages}{104017} (\bibinfo{year}{2008}), \bibinfo{note}{[Erratum:
  Phys. Rev.D79,129901(2009)]}, \eprint{0710.2335}.

\bibitem[{\citenamefont{Ajith et~al.}(2011)}]{Ajith:2009bn}
\bibinfo{author}{\bibfnamefont{P.}~\bibnamefont{Ajith}} \bibnamefont{et~al.},
  \bibinfo{journal}{Phys. Rev. Lett.} \textbf{\bibinfo{volume}{106}},
  \bibinfo{pages}{241101} (\bibinfo{year}{2011}), \eprint{0909.2867}.

\bibitem[{\citenamefont{Yunes and Pretorius}(2009)}]{PPE}
\bibinfo{author}{\bibfnamefont{N.}~\bibnamefont{Yunes}} \bibnamefont{and}
  \bibinfo{author}{\bibfnamefont{F.}~\bibnamefont{Pretorius}},
  \bibinfo{journal}{Phys.Rev.} \textbf{\bibinfo{volume}{D80}},
  \bibinfo{pages}{122003} (\bibinfo{year}{2009}).

\bibitem[{\citenamefont{Chatziioannou et~al.}(2012)\citenamefont{Chatziioannou,
  Yunes, and Cornish}}]{Chatziioannou:2012rf}
\bibinfo{author}{\bibfnamefont{K.}~\bibnamefont{Chatziioannou}},
  \bibinfo{author}{\bibfnamefont{N.}~\bibnamefont{Yunes}}, \bibnamefont{and}
  \bibinfo{author}{\bibfnamefont{N.}~\bibnamefont{Cornish}},
  \bibinfo{journal}{Phys.Rev.} \textbf{\bibinfo{volume}{D86}},
  \bibinfo{pages}{022004} (\bibinfo{year}{2012}).

\bibitem[{LIG()}]{LIGOSn}
\bibinfo{note}{{\tt{https://dcc.ligo.org/cgi-bin/DocDB/ShowDocument?docid=2974}}}.

\bibitem[{noi()}]{noise-data}
\bibinfo{note}{{\tt https://dcc.ligo.org/LIGO-T1600030/public}}.

\bibitem[{\citenamefont{Berti et~al.}(2005)\citenamefont{Berti, Buonanno, and
  Will}}]{Berti:2004bd}
\bibinfo{author}{\bibfnamefont{E.}~\bibnamefont{Berti}},
  \bibinfo{author}{\bibfnamefont{A.}~\bibnamefont{Buonanno}}, \bibnamefont{and}
  \bibinfo{author}{\bibfnamefont{C.~M.} \bibnamefont{Will}},
  \bibinfo{journal}{Phys. Rev. D} \textbf{\bibinfo{volume}{71}},
  \bibinfo{pages}{084025} (\bibinfo{year}{2005}).

\bibitem[{\citenamefont{{Heavens}}(2009)}]{2009arXiv0906.0664H}
\bibinfo{author}{\bibfnamefont{A.}~\bibnamefont{{Heavens}}},
  \bibinfo{journal}{ArXiv e-prints}  (\bibinfo{year}{2009}),
  \eprint{0906.0664}.

\bibitem[{\citenamefont{Cornish}()}]{neil}
\bibinfo{author}{\bibfnamefont{N.}~\bibnamefont{Cornish}},
  \emph{\bibinfo{title}{private communication}}.

\bibitem[{\citenamefont{Abbott
  et~al.}(2016{\natexlab{b}})}]{TheLIGOScientific:2016src}
\bibinfo{author}{\bibfnamefont{B.~P.} \bibnamefont{Abbott}}
  \bibnamefont{et~al.} (\bibinfo{collaboration}{Virgo, LIGO Scientific})
  (\bibinfo{year}{2016}{\natexlab{b}}), \eprint{1602.03841}.

\bibitem[{Sci()}]{ScienceWhite}
\emph{\bibinfo{title}{{Instrument Science White Paper}}}, \bibinfo{note}{{LIGO}
  Technical Document, {\tt https://dcc.ligo.org/LIGO-T1400316/public}}.

\bibitem[{\citenamefont{Sampson
  et~al.}(2014{\natexlab{b}})\citenamefont{Sampson, Cornish, and
  Yunes}}]{Sampson:2013jpa}
\bibinfo{author}{\bibfnamefont{L.}~\bibnamefont{Sampson}},
  \bibinfo{author}{\bibfnamefont{N.}~\bibnamefont{Cornish}}, \bibnamefont{and}
  \bibinfo{author}{\bibfnamefont{N.}~\bibnamefont{Yunes}},
  \bibinfo{journal}{Phys. Rev.} \textbf{\bibinfo{volume}{D89}},
  \bibinfo{pages}{064037} (\bibinfo{year}{2014}{\natexlab{b}}),
  \eprint{1311.4898}.

\bibitem[{\citenamefont{Cornish et~al.}(2011)\citenamefont{Cornish, Sampson,
  Yunes, and Pretorius}}]{cornish-PPE}
\bibinfo{author}{\bibfnamefont{N.}~\bibnamefont{Cornish}},
  \bibinfo{author}{\bibfnamefont{L.}~\bibnamefont{Sampson}},
  \bibinfo{author}{\bibfnamefont{N.}~\bibnamefont{Yunes}}, \bibnamefont{and}
  \bibinfo{author}{\bibfnamefont{F.}~\bibnamefont{Pretorius}},
  \bibinfo{journal}{Phys.Rev.} \textbf{\bibinfo{volume}{D84}},
  \bibinfo{pages}{062003} (\bibinfo{year}{2011}).

\bibitem[{\citenamefont{Jiménez et~al.}(2015)\citenamefont{Jiménez, Piazza,
  and Velten}}]{Jimenez:2015bwa}
\bibinfo{author}{\bibfnamefont{J.~B.} \bibnamefont{Jiménez}},
  \bibinfo{author}{\bibfnamefont{F.}~\bibnamefont{Piazza}}, \bibnamefont{and}
  \bibinfo{author}{\bibfnamefont{H.}~\bibnamefont{Velten}}
  (\bibinfo{year}{2015}), \eprint{1507.05047}.

\bibitem[{\citenamefont{de~Rham
  et~al.}(2013{\natexlab{a}})\citenamefont{de~Rham, Tolley, and
  Wesley}}]{deRham:2012fw}
\bibinfo{author}{\bibfnamefont{C.}~\bibnamefont{de~Rham}},
  \bibinfo{author}{\bibfnamefont{A.~J.} \bibnamefont{Tolley}},
  \bibnamefont{and} \bibinfo{author}{\bibfnamefont{D.~H.}
  \bibnamefont{Wesley}}, \bibinfo{journal}{Phys. Rev. D}
  \textbf{\bibinfo{volume}{87}}, \bibinfo{pages}{044025}
  (\bibinfo{year}{2013}{\natexlab{a}}).

\bibitem[{\citenamefont{de~Rham
  et~al.}(2013{\natexlab{b}})\citenamefont{de~Rham, Matas, and
  Tolley}}]{deRham:2012fg}
\bibinfo{author}{\bibfnamefont{C.}~\bibnamefont{de~Rham}},
  \bibinfo{author}{\bibfnamefont{A.}~\bibnamefont{Matas}}, \bibnamefont{and}
  \bibinfo{author}{\bibfnamefont{A.~J.} \bibnamefont{Tolley}},
  \bibinfo{journal}{Phys. Rev. D} \textbf{\bibinfo{volume}{87}},
  \bibinfo{pages}{064024} (\bibinfo{year}{2013}{\natexlab{b}}).

\bibitem[{\citenamefont{Cardoso et~al.}(2011)\citenamefont{Cardoso,
  Chakrabarti, Pani, Berti, and Gualtieri}}]{Cardoso:2011xi}
\bibinfo{author}{\bibfnamefont{V.}~\bibnamefont{Cardoso}},
  \bibinfo{author}{\bibfnamefont{S.}~\bibnamefont{Chakrabarti}},
  \bibinfo{author}{\bibfnamefont{P.}~\bibnamefont{Pani}},
  \bibinfo{author}{\bibfnamefont{E.}~\bibnamefont{Berti}}, \bibnamefont{and}
  \bibinfo{author}{\bibfnamefont{L.}~\bibnamefont{Gualtieri}},
  \bibinfo{journal}{Phys. Rev. Lett.} \textbf{\bibinfo{volume}{107}},
  \bibinfo{pages}{241101} (\bibinfo{year}{2011}), \eprint{1109.6021}.

\end{thebibliography}
\end{document}